\documentclass{article}
\usepackage{graphicx}
\usepackage{bm}
\graphicspath{{./fig/}{./png/}}
\newcommand{\EQ}{\begin{equation}}
\newcommand{\EN}{\end{equation}}
\newcommand{\EQA}{\begin{eqnarray}}
\newcommand{\ENA}{\end{eqnarray}}

\newcommand{\Eq}[1]{Eq.~(\ref{#1})}

\newcommand{\eqs}[2]{(\ref{#1}) and~(\ref{#2})}
\newcommand{\Eqss}[2]{Eqs.~(\ref{#1})--(\ref{#2})}

\newcommand{\Sec}[1]{Section~\ref{#1}}
\newcommand{\Fig}[1]{Fig.~\ref{#1}}

\newcommand{\xx}{\mbox{\boldmath $x$} {}}

\newcommand{\FF}{\mbox{\boldmath $F$} {}}

\newcommand{\grav}{\mbox{\boldmath $g$} {}}
\newcommand{\nab}{\mbox{\boldmath $\nabla$} {}}

\newcommand{\dd}{{\rm d} {}}

\def\la{\mathrel{\mathchoice {\vcenter{\offinterlineskip\halign{\hfil
$\displaystyle##$\hfil\cr<\cr\sim\cr}}}
{\vcenter{\offinterlineskip\halign{\hfil$\textstyle##$\hfil\cr<\cr\sim\cr}}}
{\vcenter{\offinterlineskip\halign{\hfil$\scriptstyle##$\hfil\cr<\cr\sim\cr}}}
{\vcenter{\offinterlineskip\halign{\hfil$\scriptscriptstyle##$\hfil\cr<\cr\sim\cr}}}}}

\newcommand{\ypnas}[5]{ (#1) #5, {\em Proc.\ Natl.\ Acad.\ Sci.\ }{\rm #2}, #3--#4.}

\newcommand{\ysci}[5]{ (#1) #5, {\em Science }{\rm #2}, #3--#4.}
\newcommand{\yicarus}[5]{ (#1) #5, {\em Icarus }{\rm #2}, #3--#4.}

\newcommand{\ynat}[5]{ (#1) #5, {\em Nature }{\rm #2}, #3--#4.}

\newcommand{\ypreN}[4]{ (#1) #4, {\em Phys.\ Rev.\ E }{\rm #2}, #3.}

\newcommand{\yptrs}[5]{ (#1) #5, {\em Phil.\ Trans.\ Roy.\ Soc.\ }{\rm #2}, #3--#4.}

\newcommand{\yjgr}[5]{ (#1) #5, {\em J.\ Geophys.\ Res.\ }{\rm #2}, #3--#4.}

\newcommand{\yrmp}[5]{ (#1) #5, {\em Rev.\ Mod.\ Phys.\ }{\rm #2}, #3--#4.}

\newcommand{\yphy}[5]{ (#1) #5, {\em Physica } {\rm #2}, #3--#4.}

\newcommand{\yoleb}[5]{ (#1) #5, {\em Orig.\ Life Evol.\ Biosph.\ }{\rm #2}, #3--#4.}
\newcommand{\yjas}[5]{ (#1) #5, {\em J. Atmos.\ Sci.\ }{\rm #2}, #3--#4.}
\newcommand{\yab}[5]{ (#1) #5, {\em Astrobiol.\ }{\rm #2}, #3--#4.}
\newcommand{\yabS}[5]{ (#1) #5 {\em Astrobiol.\ }{\rm #2}, #3--#4.}

\newcommand{\yija}[5]{ (#1) #5, {\em Int. J. Astrobiol.\ }{\rm #2}, #3--#4.}
\newcommand{\yijaS}[5]{ (#1) #5 {\em Int. J. Astrobiol.\ }{\rm #2} #3--#4.}
\newcommand{\yrpp}[5]{ (#1) #5, {\em Rep.\ Prog.\ Phys.\ }{\rm #2}, #3--#4.}

\newcommand{\yjour}[6]{ (#1) #6, {\em #2} {\rm #3}, #4--#5.}

\newcommand{\ybook}[3]{ (#1) {\em #2}, #3.}

\def\half{{\textstyle{1\over2}}}

\def\onethird{{\textstyle{1\over3}}}

\def\quarter{{\textstyle{1\over4}}}
\newcommand{\mW}{\,{\rm mW}}
\newcommand{\W}{\,{\rm W}}

\newcommand{\K}{\,{\rm K}}

\newcommand{\m}{\,{\rm m}}
\newcommand{\km}{\,{\rm km}}

\newcommand{\kW}{\,{\rm kW}}

\newcommand{\yr}{\,{\rm yr}}

\newcommand{\Gyr}{\,{\rm Gyr}}

\newcommand{\mol}{\,{\rm mol}}

\newcommand{\kJ}{\,{\rm kJ}}

\newcommand{\C}{\,{\rm C}}
\newcommand{\const}{\,{\rm const}}

\title{Nonlinear aspects of astrobiological research}
\author{Axel Brandenburg\\ Nordita, AlbaNova University Center,\\
Roslagstullsbacken 23, 10691 Stockholm, Sweden}
\date{\small Original version 13 August 2007; revised 1 September 2008}
\begin{document}
\maketitle

\begin{abstract}
Several aspects of mathematical astrobiology are discussed.
It is argued that around the time of the origin of life the handedness
of biomolecules must have established itself through an instability.
Possible pathways of producing a certain handedness include mechanisms
involving either autocatalysis or, alternatively, epimerization as
governing effects.
Concepts for establishing hereditary information are
discussed in terms of the theory of hypercycles.
Instabilities toward parasites and possible remedies by invoking
spatial extent are reviewed.
Finally, some effects of early life are discussed that contributed
to modifying and regulating atmosphere and climate of the Earth,
and that could have contributed to the highly oxidized state of its crust.
\end{abstract}

\section*{Glossary}

\subsection*{Chiral, achiral and racemic}

A molecule is chiral if its three-dimensional structure is different
from its mirror image.
Such molecules tend to be optically active and turn the polarization
plane of linearly polarized light in the right or left handed sense.
Correspondingly, they are referred to as {\sc D}- and {\sc L}-forms,
which stands for dextrorotatory and levorotatory molecules.
An achiral molecule is mirror symmetric and does not have this property.
A substance is racemic if it consists of equally many left and right
handed molecules.
A polymer is said to be isotactic if all its elements have the same chirality.

\subsection*{Enantiomers and enantiomeric excess}

Enantiomers denote a pair of chiral molecules that have opposite
handedness, but are otherwise identical.
Enantiomeric excess, usually abbreviated as e.e., is a normalized measure
of the degree by which one handedness dominates over the other one.
It is defined as the ratio of the difference to the sum of the
two concentrations, so e.e.\ is between $-1$ and $+1$.

\subsection*{Epimerization and racemization}

Epimerization is the spontaneous change of handedness of one sub-unit
in a polymer.
Racemization indicates the loss of a preferred handedness in
a substance.

\subsection*{Catalysis and auto-catalysis}

Catalysts are agents that lower the reaction barrier.
A molecule reacts with the catalyst, but at the end of the reaction
the catalyst emerges unchanged.
This is called catalysis.
In auto-catalysis the catalyst is a target molecule itself,
so this process leads to exponential amplification of the
concentration of this molecule by using some substrate.
Biological catalysts are referred to as enzymes.

\subsection*{Nucleotides and nucleic acids}

Nucleotides are the monomers of nucleic acids, e.g.\ of
RNA (ribonucleic acid) or DNA (deoxyribonucleic acid).
They contain one of four nucleobases (often just called bases)
that can pair in a specific way.
Nucleotides can form polymers and their sequence carries genetic information.
One speaks about a polycondensation reaction instead of polymerization
because one water molecule is removed in this step.
Other nucleotides of interest include peptide nucleic acid or PNA.
Here the backbone is made of peptides instead of sugar phosphate.

\subsection*{Peptides and amino acids}

Amino acids are molecules of the general form NH$_3$--CHR--COOH, where
R stands for the rest, which makes the difference between different
amino acids.
For glycine, the simplest amino acid, we have R=H, so two of the bonds
on the central C atom are the same and the molecule is therefore chiral.
For alanine, R=CH$_3$, so all four bonds on the central C atom are
different, so this molecule is chiral.
A peptide is a polymer generated through a polycondensation reaction of
amino acids.
Peptides are also referred to as proteins.

\subsection*{Solar constant and albedo}

The solar constant is the total flux of energy from the Sun above the
Earth atmosphere.
Its current value is $S=1.37\kW\m^{-2}$, but it has been about 30\%
lower when the solar system was young ($10^8\yr$ old, say), so $S$
is not a constant.
The albedo $A$ is the fraction that is reflected from the surface of the
Earth, e.g.\ by clouds and snow and, to a lesser extend, by land masses
and oceans.

\subsection*{Photosynthesis and carbon fixation}

Photosynthesis uses light to reduce CO$_2$ and to produce oxygen
either as free molecular oxygen or in some other chemical form.
This process removes CO$_2$ from the atmosphere and produces biomass,
which is written in simplistic form as $(\mbox{CH}_2\mbox{O})_n$.
This process is referred to as carbon fixation.

\subsection*{Life}

A preliminary definition of life involves replication and death,
coupled to a metabolism that utilizes any sort of available energy.
Life is characterized further by natural selection to adapt to
environmental changes and to utilize available niches.
A proper definition of life is difficult given that all life on Earth
can be traced back to a single common ancestor.
Any definition of life may need to be adjusted if extraterrestrial or
artificial life are discovered.

\section*{Definition of the subject}

Astrobiology is concerned with questions regarding possible origins of life
on Earth and elsewhere in the Universe.
Although there is presently no detection of extraterrestrial life,
it is generally assumed that life could be wide-spread provided
certain conditions of habitability are met.
A common implicit hypothesis in astrobiology is that life can emerge
spontaneously once certain environmental conditions are met.
This implies that there may well have been multiple geneses, separated
only by global extinction events such as major impacts by other celestial
bodies \cite{DaviesLineweaver}.

Four important discoveries can be named that have provided impetus
to the field of astrobiology.
\begin{enumerate}
\item More than 300 extrasolar planets have been discovered since 1995,
providing explicit targets for detecting life outside the solar system.
\item Recent Mars missions have provided evidence for liquid water
on the surface of Mars in past and possibly even present times.
This has fostered the search for techniques to detect microbial life
on Mars.
\item On Earth the carbon in very old sedimentary rocks dating back 3.8 Gyr
ago show a consistently lower $^{13}$C to $^{12}$C abundance ratio.
This is normally indicative of life.
This lends support to the notion that life may have been present
as soon as the Earth surface became hospitable.
\item The discovery of extremophiles on Earth has considerably
extended the definition of habitability to include extreme temperatures,
pressures and pH values, high salinity as well as high radiation levels.
This has raised hopes of finding life even in our solar system.
\end{enumerate}

Astrobiology thus comprises several scientific disciplines:
astronomy, geology, chemistry, and biology.
Therefore, much of the original literature tends to appear in journals
in the respective fields.
We should also mention that there are technological attempts in
producing artificial life \cite{Rasmussen_etal}.
While this approach is not aimed at reproducing the origin of life
on Earth, it may still be useful for feeding our imagination in
understanding the transition from nonliving to living matter.

\section{Introduction}

Since the early days of nonlinear dynamics and non-equilibrium
thermodynamics it has been clear that one of the ultimate applications
of this theory might be to facilitate an understanding of the transition
from non-living to living matter.
The main reason is obviously that living systems are very far from
equilibrium---as indicated by the high degree of order and hence the low
entropy of living systems relative to their exterior.

Already in 1952 Turing \cite{Turing} proposed the idea that
chemical reaction-diffusion systems might provide a tool for studying
biochemical pattern formation which have increased our understanding
of the laws of nature far from equilibrium, where life occurs.
This idea was followed up in the late 1960s by Prigogine
\cite{Prigogine,Prigogine2}
who suggested that dissipative structures have great importance in
establishing a physical description of living matter.
A general theory of autocatalytic molecular evolution was developed
in 1971 by Eigen \cite{Eigen71} who argued that in a single micro-environment
only a single handedness can result from a single event.
In particular the famous chicken and egg problem that occurs in biology
at different levels was identified as a Hopf bifurcation.
A Hopf bifurcation describes the spontaneous emergence of an oscillating
solution once some stability threshold has been crossed.
The mathematics of this is familiar to any physicist, but it requires
that the equations describing the relevant physics are known.
In biology it is not even clear that the various phenomena can be
described by equations.
A first detailed attempt in this direction was indeed that of Eigen.
However, the equations governing the emergence of life are only
phenomenological ones.
Nevertheless, these approaches are invaluable in that they help giving
the origin of life question a mathematical basis.

One of the earliest anticipated forms of life that is still similar to
present life is the RNA world \cite{Gilbert}, whereby simple RNA molecules
with functional behavior self-reproduces using genetic information encoded
either in the same or in other participating RNA molecules.
Obviously, there are tremendous difficulties given that RNA is too
complicated a molecule to be synthesized abiologically.
A significantly simpler molecule is peptide nucleic acid or PNA
\cite{Nilsen93}, where the backbone consists of peptide instead of
sugar phosphate.
Nevertheless, the difficulty of producing RNA remains.

There is no firm idea where on Earth such molecular replication may have
taken place.
Frequently discussed scenarios include hydrothermal vent systems
\cite{Russell06}, but also beach scenarios are discussed that are
subject to tides leading to cyclic changes in concentration \cite{Lathe}
as well as to repeated wetting and drying \cite{Bywater}.

An early experiment that contributed significantly to the research
into origins of life was the Urey--Miller experiment \cite{Miller53}
that demonstrated the spontaneous production of amino acids in a
reducing atmosphere consisting of H$_2$O vapor, CH$_4$, NH$_3$, and H$_2$
with an energy supply in the form of sparks.
More recent experiments also allow for the presence of CO$_2$, which now
seems unavoidable on the early Earth given that it is continuously
replenished through outgassing by volcanoes.

In the following we review some aspects where there has been considerable
cross-fertilization between astrobiology and nonlinear dynamics.
We begin by discussing a phenomenon that is believed to have taken place
around the time of the origin of life, namely the establishment of a
definitive handedness of biomolecules that is inherent to DNA and RNA
({\sc D}-form) and to amino acids ({\sc L}-form).
Next we discuss constraints on the evolution of hereditary information,
and finally review some models that characterize the alteration of the
terrestrial environment by early life.
In addition to any of these physical effects there are random fluctuations
leading inevitably to local imbalances between the concentrations of
molecules of {\sc D}- and {\sc L}-form.
In the following we discuss mechanisms that can lead to an exponential
amplification of the enantiomeric excess.
For a recent review of these ideas see Ref.~\cite{Plasson07}.

\section{Homochirality}

Theories of a chemical origin of life involve polymerization of
nucleotides that carry and utilize genetic information.
Ribonucleotides possess chirality, i.e.\ they are different from their
mirror image.
All known life forms use ribonucleotides of the so-called {\sc D}-form
(right-handed), as opposed to the  {\sc L}-form (left-handed).
These two molecules are referred to as opposite {\it enantiomers}.
In most cases these different enantiomers are optically active, i.e.\
they turn the polarization plane of linearly polarized light in a
right-handed or left-handed sense.

Any non-enzymatic synthesis of ribonucleotides would have produced
a mixture of equally many right- and left-handed building blocks.
Technically this is referred to as a {\it racemic} mixture of these
molecules.
However, experimentally it is known that in a racemic mixture of
mononucleotides the polymerization is quickly terminated after
the first or second polymerization step \cite{Joyce84}.
This is generally referred to as {\it enantiomeric cross-inhibition},
which was long thought to be a serious obstacle to a chemical origin
of life.
It was therefore thought to be necessary that life evolved
only in a homochiral environment.
Moreover, it would then be necessary that the degree of enantiomeric
purity must have been very high.
This is important because it rules out a number of physical mechanisms
based on the enantioselective effects of circularly polarized radiation,
magnetic fields, and the parity-breaking property of the electroweak
force.

\subsection{The Frank-mechanism}

A general mechanism for producing complete homochirality was proposed
in 1953 by Frank \cite{Frank53} based on the assumed effects of
auto-catalysis and what he called mutual antagonism.
In fact, the enantiomeric cross-inhibition mentioned above can be
thought of as a possible example of mutual antagonism.
The model of Frank is characterized by the following set of three reactions:
\EQA
\label{D+S}
D+S\stackrel{k_C}{\longrightarrow}D+D,\\
\label{L+S}
L+S\stackrel{k_C}{\longrightarrow}L+L,\\
D+L\stackrel{k_I}{\longrightarrow}DL,
\ENA
where $D$ and $L$ denote monomers of the two enantiomers,
$S$ is a substrate from which monomers could be formed via auto-catalysis,
and $DL$ are inactive dimers that are lost from the system.
(At this level of simplification no distinction between $DL$ and $LD$
is made. This simplification will later be relaxed.)
The parameters $k_C$ and $k_I$ characterize the reaction speeds.
These reactions translate to the following set of equations for the
concentrations of $D$, $L$, $S$, and $DL$,
\EQA
\label{dDdt}
{\dd\over\dd t}[D]&=&+k_C[S][D]-k_I[D][L],\\
\label{dLdt}
{\dd\over\dd t}[L]&=&+k_C[S][L]-k_I[D][L],\\
\label{dSdt}
{\dd\over\dd t}[S]&=&-k_C[S]\Big([D]+[L]\Big),\\
\label{dDLdt}
{\dd\over\dd t}[DL]&=&+2k_I[D][L].
\ENA
These equations imply that the total mass of all building blocks
(including the substrate), is constant, i.e.\ 
$[D]+[L]+[S]+[DL]=\mbox{const}\equiv M$.

\begin{figure}[t!]\begin{center}
\includegraphics[width=\columnwidth]{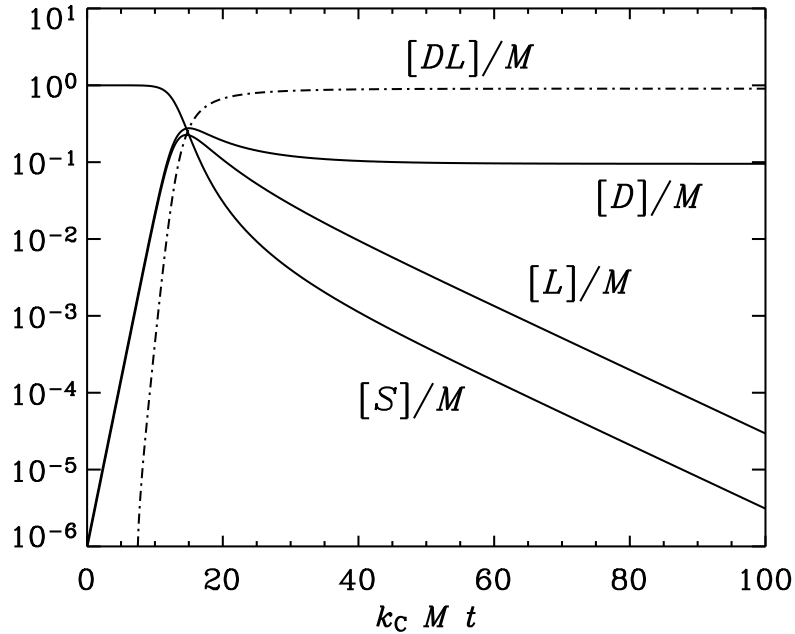}
\end{center}\caption[]{
Solution of \Eqss{dDdt}{dSdt} for $k_I=k_C$.
Both $D$ and $L$ grow exponentially until $[D]+[L]$ becomes comparable
to the constantly declining substrate concentration $[S]$.
At the same time the production of $DL$ removes an equal amount of $D$
and $L$, but this effect affects mostly those enantiomers that are
already in the minority.
In this calculation an initial asymmetry (here 10\%) of $[D]-[L]$ grows 
until saturation.
At the end,  $[D]$ has reached 100\% enantiomeric excess, but this
happened at the expense of producing a large number of inactive
heterochiral dimers $DL$.
}\label{pfrank}\end{figure}

This system of equations describes the continued autocatalytic production
of $DL$, $D$ and $L$ until the substrate $S$ is exhausted, i.e.\ $[S]=0$.
However, as long as $[S]$ is still finite, the asymmetry, ${\cal A}=[D]-[L]$,
grows quasi-exponentially, proportional to $\exp(\int[S]\,\dd t)$.
A numerical example of this is shown in \Fig{pfrank}.

In the numerical example above we started with very small initial
concentrations.
Another possibility is to start with a perturbed racemic solution.
The racemic solution is given by $[D]=[L]=\lambda/k_I$, where
$\lambda=k_C[S]$ is the instantaneous growth rate due to autocatalysis.
Under the assumption that $\lambda$ can be treated as a constant
(i.e.\ when the system is still nearly racemic),
a linear stability analysis shows that the enantiomeric excess,
\EQ
\mbox{e.e.}={[D]-[L]\over[D]+[L]}
\EN
grows exponentially.
This means that the racemic solution is unstable and that the mechanism
for achieving homochirality is based on a linear instability.

\subsection{Continued polymerization}
\label{ContinuedPolymerization}

\begin{figure}[t!]\begin{center}
\includegraphics[width=\columnwidth]{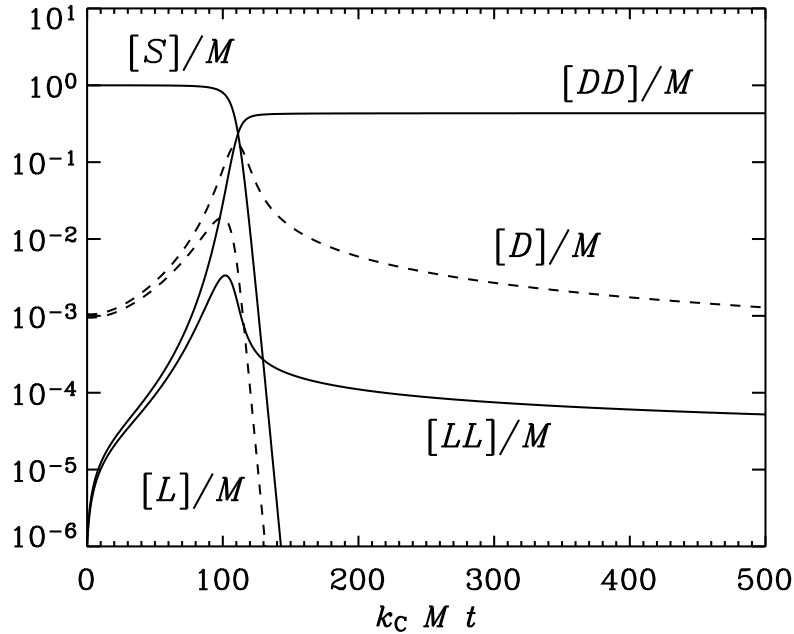}
\end{center}\caption[]{
Solution of \Eqss{dDdt}{dSdt} supplemented by the kinetic equations
corresponding to the reactions \Eqss{DDform}{LLDform},  
for $k_S=k_I=k_C$.
Again, an initial 10\% asymmetry of $[D]-[L]$ grows until $[D]+[L]$
becomes comparable to the constantly declining substrate
concentration $[S]$.
The monomers polymerize into dimers $DD$ and $LL$.
Toward the end,  $[DD]$ reaches a 100\% enantiomeric excess
}\label{pfrank2}\end{figure}

There is {\it a priori} no good reason to permit the production of
heterochiral dimers $DL$, but not of homochiral dimers $DD$ and $LL$,
i.e.\
\EQA
\label{DDform}
D+D\stackrel{k_S}{\longrightarrow}DD,\\
L+L\stackrel{k_S}{\longrightarrow}LL.
\ENA
The importance of such reactions was stressed in a review by Blackmond
\cite{Blackmond04},
who also introduced an additional modification that consists in the
assumption that not the monomers, but the homochiral dimers $DD$ and $LL$
catalyze the production of monomers, i.e.\ reactions \eqs{D+S}{L+S}
are replaced by
\EQA
DD+S\stackrel{k_C}{\longrightarrow}DD+D,\\
LL+S\stackrel{k_C}{\longrightarrow}LL+L.
\ENA
This model is similar to the original Frank model provided there is a
way of getting rid of those homochiral dimers that are in the minority.
This requires enantiomeric cross-inhibition for dimers to form heterochiral
trimers, i.e.\ we need the additional reactions
\EQA
DD+L\stackrel{k_I}{\longrightarrow}DDL,\\
\label{LLDform}
LL+D\stackrel{k_I}{\longrightarrow}LLD.
\ENA
A solution to the corresponding reaction equations is given in \Fig{pfrank2}.
Reaction calorimetry is able to give support to the assumption that
dimers and not monomers are the relevant catalysts \cite{Blackmond04}.
This seems to apply in particular to the first autocatalytic reaction ever
found that enhances enantiomeric excess \cite{Soai95}.
In this reaction (sometimes referred to as the Soai reaction) the
substrate is pyridine-3-carbaldehyde and the chiral molecule of either
{\sc D}- or {\sc L}-form is 3-pyridyl alkanol which thus acts as an
asymmetric autocatalyst to produce more of itself.
In this reaction, however, dialkylzinc acts as an additional achiral catalysts.
While the Soai reaction is important as a first explicit example of an
autocatalytic reaction that enhances the enantiomeric excess, it is not
normally regarded as directly important for astrobiology.

The polymerization model has been developed by Sandars \cite{Sandars03},
who included arbitrarily many polymerization steps of the form
\EQA
D_n+D\stackrel{k_S}{\longrightarrow}D_{n+1},\\
L_n+L\stackrel{k_S}{\longrightarrow}L_{n+1},\\
D_n+L\stackrel{k_S}{\longrightarrow}D_nL,\\
L_n+D\stackrel{k_S}{\longrightarrow}L_nD.
\ENA
The basic outcome of this and similar models is always the same
as in the original Frank model, except that the polymerization model
is also capable of displaying interesting wave-like dynamics in
time-dependent histograms of different polymers \cite{BAHN05}.

\subsection{Spatially extended models}

In reality there are limits as to how much a system can be considered
fully mixed.
In general, $[D]$ and $[L]$ should be functions of time {\it and}
space, i.e.\ $[D]=[D](t,\xx)$ and $[L]=[L](t,\xx)$.
Assuming that there is only molecular diffusion, the relevant reaction
equations are to be supplemented by additional diffusion terms,
\EQA
\label{dDdtdiff}
{\dd\over\dd t}[D_n]=R_n^{(D)}+\kappa\nabla^2[D_n],\\
\label{dLdtdiff}
{\dd\over\dd t}[L_n]=R_n^{(L)}+\kappa\nabla^2[L_n],
\ENA
where $R_n^{(D)}$ and $R_n^{(L)}L$ are the right hand sides of the
reaction equations.

If there were only one type of handedness, the resulting equation
would be reminiscent of the Fisher equation \cite{Murray93},
\EQ
{\dd f\over\dd t}=\lambda (1-f)f+\kappa\nabla^2f,
\EN
which admits propagating front solutions with front speed
$v_{\rm front}=2\sqrt{\kappa\lambda}$.
Here, $f$ could represent the local concentration of some
disease in models of the spread of epidemics, for example.

\begin{figure}[t!]\begin{center}
\includegraphics[width=.7\columnwidth]{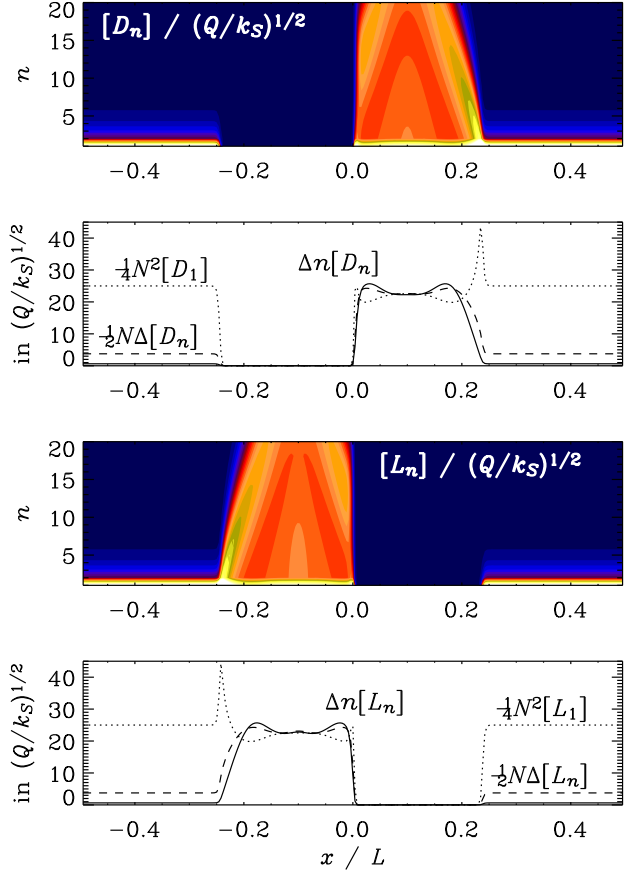}
\end{center}\caption[]{
Color/gray scale plots of $[D_n]$ and $[L_n]$ for $t/\tau_{\rm diff}=0.8$
as a function of $x$ and $n$,
and the corresponding dependencies of
$\sum_{n=1}^Nn[D_n]$ and $\sum_{n=1}^Nn[L_n]$ (solid line), compared with
$\half N\sum_{n=1}^N[D_n]$ and $\half N\sum_{n=1}^N[L_n]$ (dashed line), and
$\quarter N^2[D_1]$ and $\quarter N^2[L_1]$ (dotted line), all in units
of $(Q/k_S)^{1/2}$.
The normalized diffusivity is $\kappa/(L^2\lambda_0)=10^{-2}$
and $N=20$.
Adapted from Ref.~\cite{Brand+Mult04}.
}\label{pchain}\end{figure}

In the present case there are two fields, one of each handedness.
It is instructive to refer to these fields as populations which is
suggestive of their ability to replicate, migrate. become extinct,
and to compete against a population of opposite handedness.
Each population is able to expand into unpopulated space at a speed
given approximately by $v_{\rm front}$, but once two opposing handednesses come
into contact, there is an impasse and the propagation comes to a halt.
A snapshot of a one-dimensional model illustrating the polymer length
as a function of position is shown in \Fig{pchain} for populations of
opposite handedness that have come into contact.

\begin{figure}[t!]\begin{center}
\includegraphics[width=.7\columnwidth]{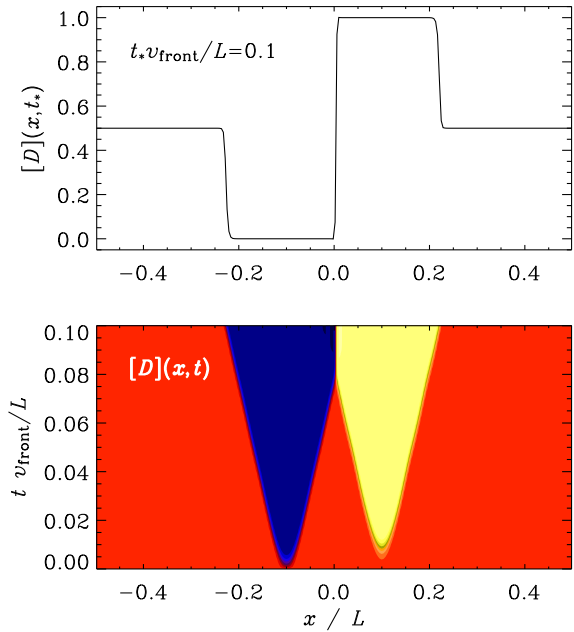}
\end{center}\caption[]{
Profile of $[D](x,t_*)$ and
space-time diagram of $[D](x,t)$ for the one-dimensional problem without
advection and an initial perturbation corresponding to a weak
(amplitude 0.01) right-handed
excess at $x/L=0.1$ (marked in white or yellow) and a somewhat stronger
(amplitude 0.3) left-handed excess at $x/L=-0.1$ (marked in dark or blue).
Note the propagation of fronts with constant speed if the exterior is
racemic (i.e.\ $[D]=[L]=1/2$, shown in medium shades or red)
and a nonpropagating front when
the chirality is opposite on the two sides of the front.
The normalized diffusivity is $\kappa/(L^2\lambda_0)=10^{-2}$,
i.e.\ the same as in \Fig{pchain}.
Adapted from Ref.~\cite{Brand+Mult04}.
}\label{p1drun}\end{figure}

The overall dynamics of symmetry breaking is well characterized by a low
order truncation, where the model
is truncated at $n=2$ and the evolution of the $n=1$ modes is assumed
to be enslaved by the evolution of the $n=2$ modes \cite{BAHN05}.
An example of such a solution is shown in \Fig{p1drun}, which shows
the evolution in a space-time diagram, where two populations of opposite
handedness expand into unpopulated space until two opposite populations
come into contact.

In two and three dimensions a front between two opposing enantiomers is
in general curved, in which case it can propagate diffusively in the
direction of curvature.
This is caused by the fact that the inner front between two populations
is slightly shorter than the outer one.
(Only the immediate proximity of a front matters; what lies behind it
is irrelevant if it is of the same handedness.)
Indeed, on a two-dimensional surface the inner front is by $2\pi d$ shorter
than the outer one---independent of radius.
Here, $d$ is the front thickness, which is of the order of
$d\approx(\kappa/\lambda)^{1/2}$.

It turns out that in two dimensions the rate of change
of the integrated asymmetry, ${\cal A}=\int([D]-[L])\,\dd^2x$, depends
only on the {\it number} of topologically distinct rings or islands.
Once an island is wiped out, the rate of change of ${\cal A}$ changes
abruptly and stays then constant until the next island gets wiped out.
So the enantiomeric excess,
\EQ
\mbox{e.e.}={\int([D]-[L])\,\dd^2x\over\int([D]+[L])\,\dd^2x},
\EN
increases with time in a piecewise linear fashion.

Even if at each point homochirality could be reached rapidly (time scale
$\lambda^{-1}$), global homochirality requires that one population wipes
out the other one completely.
Diffusion is usually too slow to lead to any significant mixing and
hence to global homochirality.
However, there could be circumstances where such mixing is sped up by
something like ``turbulent'' transport.
In the case of the Earth the slowest relevant transport is in the
Earth mantle, part of which is now associated with what is called the
deep biosphere.
Assuming that multiple geneses of life is possible, this would
raise the question whether a simultaneous co-existence of different
handednesses on different parts of the early Earth would have been possible.
It is however unclear whether this possibility could have left any
traces that would still be detectable today.

\begin{figure}[t!]
\resizebox{\hsize}{!}{\includegraphics{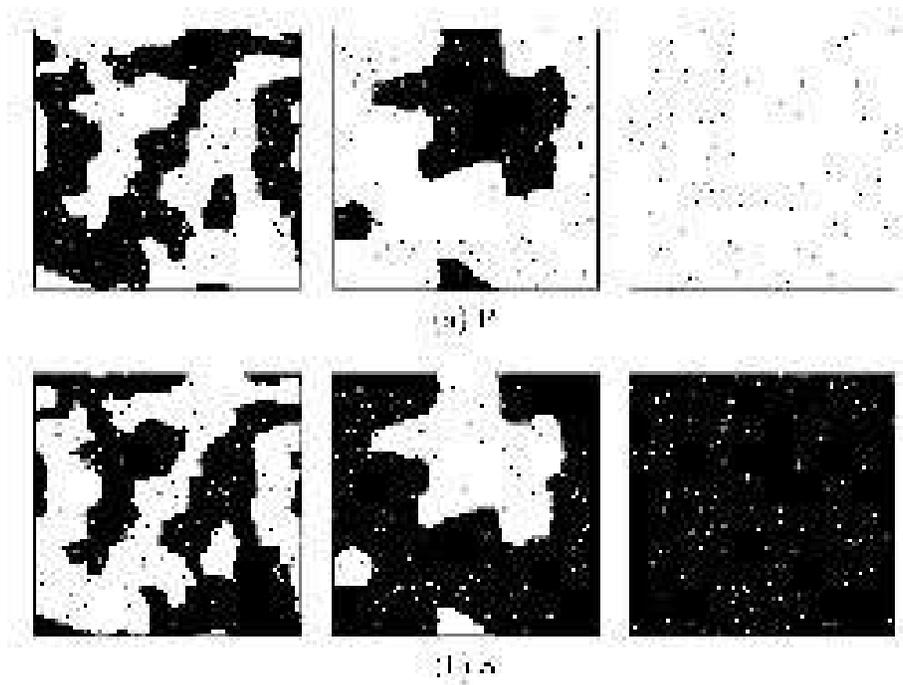}}\caption{
The evolution of molecules of {\sc D}- and {\sc L}-forms is shown in
the upper and lower panels, respectively.
Note the tendency toward complete homochiralization by gradually filling
up isolated islands with the chirality of the surrounding molecules.
The dimensionless times are 50, 250, and 1750 from left to right.
Courtesy of Shibata et al.\ \cite{Shibata06}.
}\label{shibata_etal06}\end{figure}

Another approach to solving the problem of spatially extended chemistry is
by means of cellular automata.
In this approach points on a mesh can take different states corresponding
to molecules of right or left handedness, achiral substrate molecules,
or even empty states.
An example of such a calculation by Shibata et al.\ \cite{Shibata06}
is shown in \Fig{shibata_etal06}.
Again, there are patches of populations of opposite handedness that grow
and wipe each other out such that in the end only one handedness survives.

\subsection{Epimerization}
\label{Epimerization}

\begin{figure}[t!]\begin{center}
\includegraphics[width=.7\columnwidth]{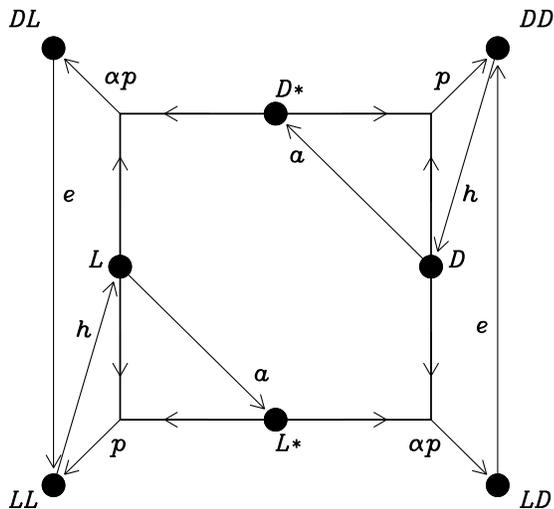}
\end{center}\caption[]{
Representation of the minimal set of reactions necessary for
allowing the spontaneous transition to homochirality.
Adapted from Ref.~\cite{BLL07}.
}\label{psketch_recycled}\end{figure}

An interesting alternative to the Frank-type mechanism is a set of
reactions based primarily on a phenomenon called epimerization, i.e.\
the spontaneous change of handedness in one part of the polymer.
This mechanism is important in the chemistry of amino acids.
Plasson et al.\ \cite{Plasson04} identified four reactions: activation,
polymerization, epimerization, and depolymerization as
necessary ingredients that can, under certain conditions, lead to
an instability of the racemic state with a bifurcation toward full
homochirality.
They called this the APED model, whose reactions can be summarized
as follows:\\
{\bf A}: activation:
\EQ
L\stackrel{a}{\longrightarrow}L^*,\quad
D\stackrel{a}{\longrightarrow}D^*,
\EN
{\bf P}: polymerization:
\EQ
L^*+L\stackrel{p}{\longrightarrow}LL,\quad
D^*+D\stackrel{p}{\longrightarrow}DD,
\EN
\EQ
L^*+D\stackrel{\alpha p}{\longrightarrow}LD,\quad
D^*+L\stackrel{\alpha p}{\longrightarrow}DL,
\EN
{\bf E}: epimerization:
\EQ
LD\stackrel{e}{\longrightarrow}DD,\quad
DL\stackrel{e}{\longrightarrow}LL,
\EN
{\bf D}: depolymerization:
\EQ
LL\stackrel{h}{\longrightarrow}L+L,\quad
DD\stackrel{h}{\longrightarrow}D+D.
\EN
This minimal subset of reactions is shown in \Fig{psketch_recycled}.

Compared with the Frank model, a major advantage of the present
model is that no hypothetical auto-catalysis is required.
Indeed, all these reactions exist in principle, although it is as yet
unclear what kind of manipulations on the environment are required to make
all these reactions happen.
Another advantage is that the system is closed, so no
inflow or outflow of matter is required.
The system is maintained away from equilibrium by energy input though
the activation of amino acids.

Given that there is neither auto-catalysis nor enantiomeric
cross-inhibition, one wonders whether the APED model still shares some
similarities with Frank's original model.
Some degree of similarity is immediately seen by writing the APED reactions
in sequential form in one line, i.e.,
\EQ
D^*+L
\stackrel{\alpha p}{\longrightarrow}DL
\stackrel{e}{\longrightarrow}LL
\stackrel{h}{\longrightarrow}L+L,
\EN
\EQ
L^*+D
\stackrel{\alpha p}{\longrightarrow}LD
\stackrel{e}{\longrightarrow}DD
\stackrel{h}{\longrightarrow}D+D.
\EN
This shows that, as long as the reaction rates for epimerization and
depolymerization are not limiting factors, we have essentially the
reactions
\EQ
D^*\stackrel{\alpha p L}{\longrightarrow}L\quad\mbox{and}\quad
L^*\stackrel{\alpha p D}{\longrightarrow}D.
\EN
This way of writing these reactions emphasizes the roles of $L$ and $D$ in
catalyzing the conversion of $D^*$ into $L$ and $L^*$ into $D$, respectively.
Just like the mechanism of mutual antagonism, these reactions disfavor
a racemic state, but instead of producing unreactive waste, these reactions
produce directly one of two possible homochiral states.

In addition, there are reactions of the form
\EQ
\label{L2L}
L^*+L
\stackrel{p}{\longrightarrow}LL
\stackrel{h}{\longrightarrow}L+L,
\EN
\EQ
D^*+D
\stackrel{p}{\longrightarrow}DD
\stackrel{h}{\longrightarrow}D+D.
\EN
These reactions simulate the autocatalytic
conversion of $L^*$ into $L$ by $L$  and of $D^*$ into $D$ by $D$.
Again, linear analysis establishes that the racemic state is unstable
provided $\alpha$ is in the range $0<\alpha<1$;
see Refs.~\cite{Plasson04,BLL07}.

In conclusion we can say that the homochirality of life-bearing
molecules might well have originated from the chemical reactions
that lead to their formation.
Thus, the hypothetical RNA world may have been born into an environment
surrounded by homochiral peptides (as described in \Sec{Epimerization}),
or, alternatively, homochirality
may have emerged as a consequence of enantiomeric cross-inhibition
during the first stages of the RNA world (as discussed in
\Sec{ContinuedPolymerization}).

In the next section we discuss some issues regarding possible
strategies for establishing a primitive information-carrying system.
This is also based on catalysis, but catalysis in the production of
other molecules than itself.

\section{Establishing hereditary information}

We have so far ignored the fact that polymers can consist of
different amino acid or nucleotide units, even though they
would all have the same handedness.
Therefore such molecules could in principle carry information.
Once such polymers can replicate,
the question arises how to prevent them from getting extinct
due to errors in the copying process, and instead to compete
against parasites.
It is generally believed that early self-replicating systems had a
substantial error rate associated with each replication event.
A certain small error rate is obviously necessary for facilitating
Darwinian evolution by natural selection,
but it must be small enough to prevent extinction.

\begin{figure}[t!]\begin{center}
\includegraphics[width=.7\columnwidth]{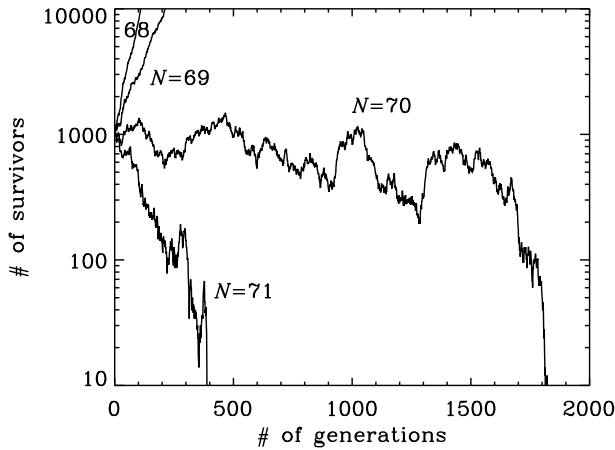}
\end{center}\caption[]{
Number of survivors as a function of the number of generations in
a numerical example with an error rate of $p=0.01$.
This initial number of survivors is 1000.
Note that for a genome length of $N=70$ and 71 the population dies
out after 1800 and 400 generations, respectively.
For $N=69$ and less the number of survivors increases exponentially.
This is compatible with the survival criterion $N\la(\ln 2)/0.01=69.3$,
derived from \Eq{GeneLength}.
}\label{pexp12}\end{figure}

Assuming that with each generation a species produces $\sigma$ offspring
where the length of the genome is $N$ bits, and that the
probability for a copying error at any position in the genome is $p$,
the necessary condition for long-term survival is given by
\cite{Eigen71,Eigen02}
\EQ
pN < \ln \sigma.
\label{GeneLength}
\EN
The significance of this formula is illustrated in \Fig{pexp12} with
the help of a numerical example where the selective advantage
(i.e.\ the multiplication factor) is chosen to be $\sigma=2$,
the error rate is $p=0.01$, and four different values of $N$
between 68 and 71 are used.
In this numerical experiment, $\sigma$ new offspring are produced, but with a
probability $p$ an error is introduced at each of the $N$ positions.
After this only the intact copies can produce further offspring,
and so forth.

According to \Eq{GeneLength} the maximum genome length is,
with the parameters of our example, $(\ln 2)/0.01=69.3$.
This is compatible with \Fig{pexp12} which shows that the
dividing line between extinction and long-term survival
is between $N=69$ and 70.
For contemporary genomes $N$ is of the order of $10^8$, and $p$ is of
the order of $10^{-8}$ \cite{Dyson88}, or below, depending on the efficiency
of error-correcting mechanisms that are available in contemporary organisms.

\begin{figure}[t!]\begin{center}
\includegraphics[width=.7\columnwidth]{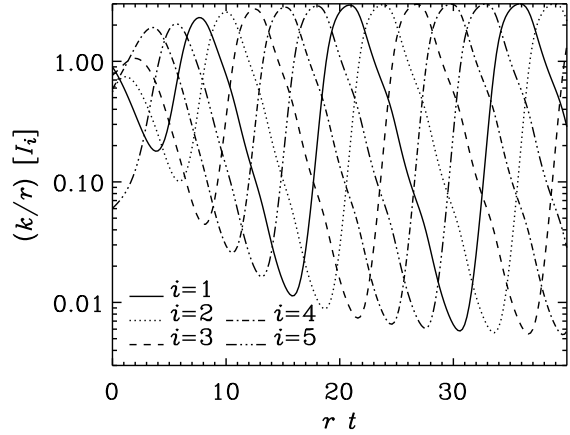}
\end{center}\caption[]{
Evolution of the number densities of five hypercycles with
equal parameters.
Note that peaks of $I_1$ (solid line) are followed by peaks of
$I_2$ (dotted line) and $I_3$ (dashed line), and so forth.
Time is measured in units of $r^{-1}$ and concentrations are
measured in units of $r/k$.
}\label{phypercyc}\end{figure}

The first replicating systems are likely to have rather high error
rates, and no correction mechanism, making it virtually impossible to
carry sufficient information for building more complex replicators.
This difficulty can be removed by invoking the concept of hypercycles
\cite{EigenSchuster77}, whereby the full genetic information is
carried collectively by several smaller systems (smaller $N$),
each one small enough to obey \Eq{GeneLength}.
Mathematically, such a system can be described by the following set
of reactions \cite{Boerlijst91}:
\EQ
I_i\stackrel{r_i}{\longrightarrow}2I_i,
\EN
\EQ
I_i+I_{i-1}\stackrel{k_i}{\longrightarrow}2I_i+I_{i-1}.
\EN
Assuming furthermore that resources are limited, the total number
of molecules, $M=\sum_i[I_i]$, is taken to be constant, i.e.\
$I_i$ is assumed to be siphoned off from the system at a rate $\phi$
that is independent of $i$.
Mathematically, such a system can be described by the following set
of ordinary differential equations:
\EQ
{\dd\over\dd t}[I_i]=r_i[I_i]+k_i[I_i][I_{i-1}]-\phi[I_i],
\EN
where
\EQ
\phi=\sum_i\Big(r_i[I_i]+k_i[I_i][I_{i-1}]\Big)\left/\sum_i[I_i]\right.
\EN
is the factor that keeps the total number of molecules constant.
The kinetic coefficient $r_i$ models the residual effects of birth and death,
while $k_i$ is the kinetic coefficient for the catalytic production of $I_i$,
where $I_{i-1}$ acts as a catalyst.
The evolution of number densities in a model of five hypercycles
is shown in \Fig{phypercyc} for a case where all $k_i=k$ and
$r_i=r$ are chosen to be the same for all values of $i$.

\begin{figure}[t!]\begin{center}
\includegraphics[width=.7\columnwidth]{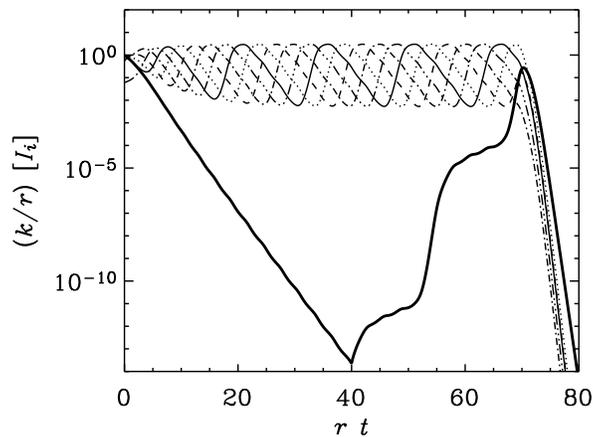}
\end{center}\caption[]{
Evolution of the number densities of five hypercycles with
equal parameters and a parasite where $k_{\rm para}=2k_i$
and $r_{\rm para}=r$.
Time is measured in units of $r^{-1}$ and concentrations are
measured in units of $r/k$.
}\label{phypercyc_para}\end{figure}

An interesting situation arises when the effects of parasites are
included.
Boerlijst \& Hogeweg \cite{Boerlijst91} considered an example where a parasite
is coupled to $I_2$; see \Eq{BH91para}.
The effect on the above model is shown in \Fig{phypercyc_para}
where $k_{\rm para}=2k$ and $r_{\rm para}=r$ has been chosen.
One sees that not much happens for a long time when the parasite
is turned on.
This is because the parasite has to grow to a level where it can
affect the entire system.
When this point is reached, all components of the system decay
exponentially---including the parasite itself.
Unfortunately, the system can never recover from this disaster,
so the hypercycle theory seems to have a problem.

\begin{figure}[t!]\begin{center}
\includegraphics[width=.7\columnwidth]{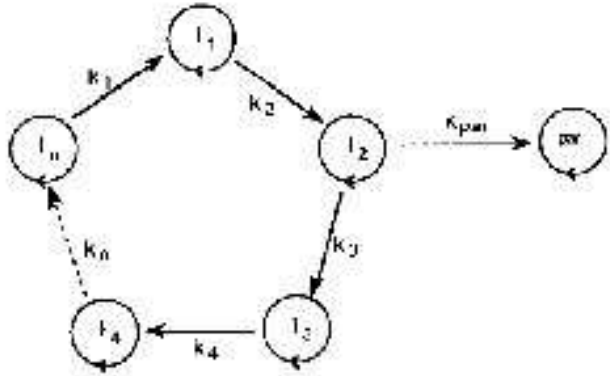}
\end{center}\caption[]{
Sketch showing the coupling of several hypercycles
together with a parasite coupled to species $I_2$.
Courtesy of M.C.\ Boerlijst \& P.\ Hogeweg \cite{Boerlijst91}.
}\label{BH91para}\end{figure}

\begin{figure}[t!]\begin{center}
\includegraphics[width=.7\columnwidth]{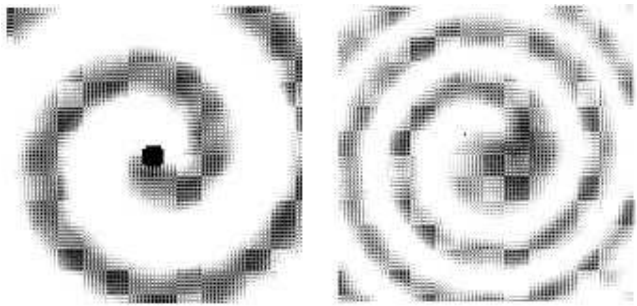}
\end{center}\caption[]{
Spatial patterns of the limit cycle showing spiral waves
around an initial obstacle (left) that was removed at later times (right).
Courtesy of M.C.\ Boerlijst \& P.\ Hogeweg \cite{Boerlijst95}.
}\label{boerlijst95}\end{figure}

Again, spatial extent can significantly alter the situation.
Using a cellular automata approach, Boerlijst \& Hogeweg \cite{Boerlijst91}
showed that the danger of parasitic catastrophes can be eliminated
by allowing the offspring to enter unpolluted areas faster than
the growth of the parasite allows.
Interestingly enough, this approach tends to produce spiraling
interfaces between different species; see \Fig{boerlijst95}.

These equations are in their nature similar to other chemical
reaction-diffusion equations where several different substances
catalyze each others reactions.
A particularly exciting example is the famous Belousov-Zhabotinsky
reaction, where malonic acid, CH$_2$(COOH)$_2$ is oxidized in the
presence of bromate ions, BrO$_3^-$.
To initiate the reaction, cerium is used as catalyst to donate ions,
although other metal ions are also possible.
The color depends on the state of the cerium as it changes from Ce$^{3+}$
to Ce$^{4+}$ or, if iron is used, from Fe$^{2+}$ to Fe$^{3+}$.
The resulting reactions are of the form \cite{Murray74}
\EQA
A+Y\stackrel{k_1}{\longrightarrow}X,\\
X+Y\stackrel{k_2}{\longrightarrow}P,\\
B+X\stackrel{k_3}{\longrightarrow}2X+Z,\\
2X\stackrel{k_4}{\longrightarrow}Q,\\
Z\stackrel{k_5}{\longrightarrow}Y,
\ENA
where $X$=HBrO$_2$, $Y$=Br$^-$, $Z$=Ce$^{4+}$, $A=B$=BrO$_3^-$,
$P$ and $Q$ are reaction products that do not contribute further to
the reactions, and $k_1, ..., k_5$ are known rate constants.
The reactions above lead to kinetic equations of the form
\EQ
{\partial[X]\over\partial t}=k_1[A][Y]-k_2[X][Y]+k_3[A][X]-k_4[X]^2,
\EN
\EQ
{\partial[Y]\over\partial t}=-k_1[A][Y]-k_2[X][Y]+k_5[Z],
\EN
\EQ
{\partial[Z]\over\partial t}=2k_3[A][X]-k_5[Z].
\EN
This model of reaction equations is called the
Oregonator, which refers to the affiliation of the authors at the
time of publication \cite{Oregonator}.

\begin{figure}[t!]\begin{center}
\includegraphics[width=.7\columnwidth]{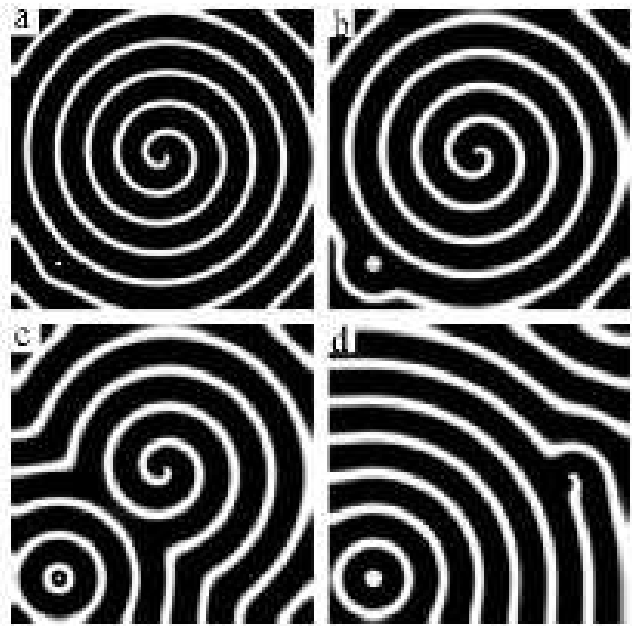}
\end{center}\caption[]{
Spiral and ring-like patterns for the modified reaction equations by
Zhang et al.\ \cite{Zhang_etal04}.
On the boundaries a no-flux condition has been adopted, i.e.\ the
normal components of all gradients vanish.
Courtesy of X.Z.\ Zhang et al.\ \cite{Zhang_etal04}.
}\label{zhang_etal04}\end{figure}

If spatial extent is included via diffusion terms, this reaction exhibits
in certain cases spiral patterns, similar to those in the model of
Boerlijst \& Hogeweg \cite{Boerlijst91}.
In \Fig{zhang_etal04} we reproduce the pattern obtained by Zhang et al.\
\cite{Zhang_etal04} for a slightly modified model consisting of only
two partial differential equations.
Depending on the value of a certain control parameter in their model
spiral patterns of different size are being produced.
An extensive review of the physics of pattern formation in different
settings is given by Cross and Hohenberg \cite{CrossHohenberg}.

The connection between pattern formation and the origin of life may seem
rather remote.
However, the equations governing chemical pattern formation illustrate
some of the critical steps that are expected to play a role in origins
of life.
In particular the fact that different chemical compounds catalyze each
other in a productive manner is an essential property behind the model
proposed by Eigen.
The Belousov-Zhabotinsky reaction also illustrates the phenomenon of
autocatalysis where, in the presence of $A$, the molecule $X$ catalyzes
the production of more $X$ by using $A$ as a substrate and producing $Z$
as additional side product.

The possibility of self-replication has been demonstrated for simple
RNA molecules by Spiegelman \cite{Spiegelman} back in the late 1960s.
There are now even examples of simple peptide chains that can catalyze
the production of each other \cite{Ghadiri}.
However, a serious shortcoming of any of the above examples discussed here
is the fact that there is no possibility of natural selection and hence
Darwinian evolution.
So, as far as the question of the origin of life is concerned this pathway
must be considered a dead end.

In summary one can say that there are similarities in the mathematics
of producing homochirality and in establishing hereditary information
in the composition of the first replicating polymers.
However, in the latter case even less is known about the detailed
nature of such polymers and their catalytic properties.
A particularly important aspect is the possibility of spatial
extent that can substantially modify the behavior of any chemical system.
In the present case, as shown in Ref.~\cite{Boerlijst91}, the possibility
of spatial extent is critical for stabilizing the system against destruction
by parasites.
The model also exhibits spiral pattern formation that has been at the
heart of the early work by Prigogine and others in connection with
early ideas on biogenesis.

\section{Alteration of the environment by early life}

In this last section we discuss some physics problems within astrobiology
that illustrate how life, once it has formed, might affect the environment
of the early Earth and how it led to a planet so
markedly different from a planet that does not harbor life.

\subsection{Global energy balance of the Earth}

The young Sun was about 30\% fainter than today, and yet the young Earth
was covered with liquid water and had temperatures higher than nowadays.
This was caused by the presence of greenhouse gases such as water vapor,
carbon dioxide, and probably also methane.
Life is responsible for reducing $\mbox{CO}_2$ to compounds of the form
$(\mbox{CH}_2\mbox{O})_n$ and similar, and to oxidize various minerals
or to produce $\mbox{O}_2$.
The resulting decrease of $\mbox{CO}_2$ weakens the greenhouse effect,
so in this sense the emergence of life has essentially a cooling effect
on the Earth's overall climate.

Without atmosphere, the planet would cool like a black body
at a rate proportional to the local flux $\sigma_{\rm SB}T^4$,
where $\sigma_{\rm SB}$ is the Stefan-Boltzmann constant.
Integrated over the entire surface of the planet, this corresponds
to a loss of $4\pi R_{\rm E}^2\sigma_{\rm SB}T^4$, which would need
to be balanced against the rate of energy received by solar radiation.
The solar ``constant'' is $S=1.37\kW\m^{-2}$ and the total energy projected
onto the disk of the Earth is $(1-A)\pi R_{\rm E}^2S$, where $A$ is the
albedo, i.e.\ the fraction of energy reflected from the Earth.
The resulting blackbody temperature would be
\EQ
T_0=\left[(1-A){S\over4\sigma_{\rm SB}}\right]^{1/4}.
\EN
Using $A=0.3$ and $\sigma_{\rm SB}=5.67\times10^{-8}\W\m^{-2}\K^{-1})$,
the temperature of the Earth would be $255\K$ or about $-18\C$.

In the presence of an atmosphere the rate of cooling is modified to
$\sigma_{\rm SB}T_{\rm eff}^4$, where $T_{\rm eff}$ is the effective
temperature equivalent to that of a black body.
A positive greenhouse effect corresponds to $T_{\rm eff}<T$, so the
cooling is reduced and the atmosphere heats up according to the
vertically integrated energy equation
\EQ
C{\dd T_0\over\dd t}=(1-A){S\over4}
-\sigma_{\rm SB}T_{\rm eff}^4,
\label{dTdt}
\EN
where $C$ is the vertically integrated specific heat.

The value of the effective temperature can be obtained from a
radiative transfer calculation.
A simplified model calculation\footnote{Under the assumption of local
isotropy (Eddington approximation) radiative equilibrium implies that
the flux is proportional to the negative gradient of the radiative
energy density $aT^4$, where $a$ is the radiation-density constant, so
$$
\FF=-\onethird c\ell\nab(aT^4).
$$
Here, $c$ is the speed of light and $\ell$ is the mean free path of photons.
The latter can be expressed in terms of the opacity $\kappa$ and the
density $\rho$ as $\ell=(\kappa\rho)^{-1}$.
Hydrostatic equilibrium can be written in the form
$$
\grav=-{{\cal R}\over\mu}\nab T,
$$
where $\grav$ is the gravitational acceleration, ${\cal R}$ is the
universal gas constant, and $\mu$ is the mean molecular weight.
These equations can be solved by a polytrope, i.e.\ $T=T_0(1-z/H)$
and $\rho=\rho_0(1-z/H)^3$, where $z$ is the distance from the surface
and $H$ is the vertical pressure scale height.
This leads to a condition of the form \Eq{TvsTeff} where
$\ell_{\rm crit}=3H/16\approx0.19H$ is the critical mean free path of photons.
} yields
\EQ
T_{\rm eff}^4={\ell\over\ell_{\rm crit}}\,T_0^4,
\label{TvsTeff}
\EN
where $T_0$ is the surface temperature, $\ell$ is an averaged mean free
path of photons and $\ell_{\rm crit}$ is the critical value above
which there is a positive greenhouse effect.
Again, a simplified calculation suggests $\ell_{\rm crit}=3H/16\approx0.19H$,
where $H={\cal R}T/(\mu g)\approx8\km$ is the pressure scale height of the
atmosphere.
So, an increase in opacity leads to a decrease of the cooling
and hence to an increase in the surface temperature.

Another interpretation is to say that the greenhouse gases shift the
radiating surface by a certain amount, $\ell_{\rm g}$, upward.
The value of $\ell_{\rm g}$ is related to $\ell$.
Ditlevsen \cite{Ditlevsen05} uses $\ell_{\rm g}=3\km$.
He also noted that a more accurate lapse rate of the temperature is
$\dd T/\dd z=10\K\km^{-1}$ instead of $T_0/H=40\K\km^{-1}$,
so that the temperature gain caused by
greenhouse gases is $\ell_{\rm g}\times\dd T/\dd z=30\K$.
The reason for a shallower temperature gradient is the
presence of convection that causes the specific entropy $s$ to be nearly
constant with height.
In that case the temperature gradient is just the adiabatic one,
$(\dd T/\dd z)_{\rm ad}=g/c_p$, where $c_p$ is the specific heat
at fixed pressure, which in turn is related to the universal gas constant
and the specific weight via ${\cal R}/\mu=c_p-c_v$, where $c_v$ is
the specific heat at fixed volume and $c_p/c_v=\gamma$ is the ratio
of specific heats.
With these formulae one does indeed get\footnote{Hydrostatic equilibrium can
be written as $-\rho^{-1}\nab p-\nab\phi=0$, where $p$ is the pressure
and $\phi=gz+\const$ is the gravitational potential.
Using the thermodynamic relation $-\rho^{-1}\nab p=-\nab h+T\nab s=0$,
where $h=c_pT$ is the specific enthalpy and $\nab s=0$ for adiabatic
stratification, we have $\dd(c_pT)/\dd z=g$.}
\EQ
\left({\dd T\over\dd z}\right)_{\rm ad}=\left(1-{1\over\gamma}\right)
{\mu g\over{\cal R}}\approx10\K\km^{-1},
\EN
where we have used $\gamma=7/5$ for air molecules with 5 degrees of
freedom (3 for translation and 2 for rotation).

At certain times over the history of the Earth other greenhouse gases
such as methane may have played an important role in keeping the Earth
above freezing temperatures.
Indeed, the burial of oxides in the crust allowed methane to build
up in the atmosphere, which may have led to concentrations of a few
thousand times greater than modern levels.
UV radiation in the upper atmosphere breaks up methane into
its components, letting H$_2$ to escape into space, leading to a
net gain of oxygen, that comes ultimately from H$_2$O.

According to a model of Catling et al.\ \cite{Catling_etal01} methane
(CH$_4$) may have been important $2.7...2.3\Gyr$ ago, just before
the famous Snowball Earth deep freeze of the Earth \cite{Hoffman}.
As discussed above, the associated loss of hydrogen may have led to a
gradual accumulation of oxygen in the atmosphere, which then terminated
the methane era and led to the Snowball Earth event.
This event lasted until the continuous CO$_2$ production from volcanoes
accumulated to large amounts so that the resulting greenhouse effect
became sufficient to initiate partial melting of the ice cover.

\subsection{Response to changes in greenhouse gases}

\begin{figure}[t!]\begin{center}
\includegraphics[width=.9\columnwidth]{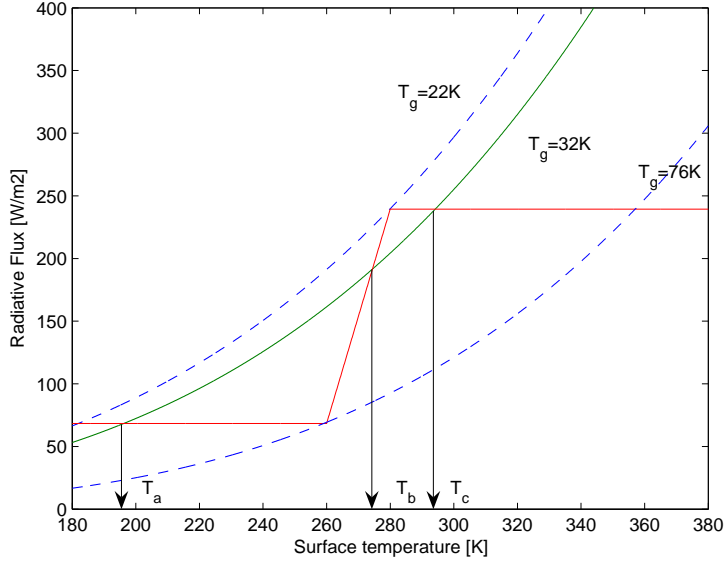}
\end{center}\caption[]{
Plot of $\sigma_{\rm SB}T_{\rm eff}^4$ versus surface temperature $T_0$
for three different greenhouse temperature shifts, $T_{\rm g}$, compared
with the net radiation ${\textstyle{1\over4}}(1-A)$ for a simple
piecewise linear function $A(T)$.
Courtesy of P.\ Ditlevsen \cite{Ditlevsen05}.
}\label{ditlevsen1}\end{figure}

As was known from global climate models \cite{Ghil} and later from
simplified models \cite{Kallen} using \Eq{dTdt} with a relatively
simple piecewise linear temperature dependence of $A(T)$, there
can be three different equilibrium temperatures.
This is illustrated in \Fig{ditlevsen1}, where we compare the graph
of $\sigma_{\rm SB}T_{\rm eff}^4$ versus surface temperature $T_0$
with the net radiation $(1-A)S/4$.
Here, $A(T)$ has been arranged such that
$A=A_{\rm hot}$ for $T\ge T_{\max}$ (corresponding to no ice coverage)
$A=A_{\rm cold}$ for $T\le T_{\min}$ (corresponding to full ice coverage).

\begin{figure}[t!]\begin{center}
\includegraphics[width=.9\columnwidth]{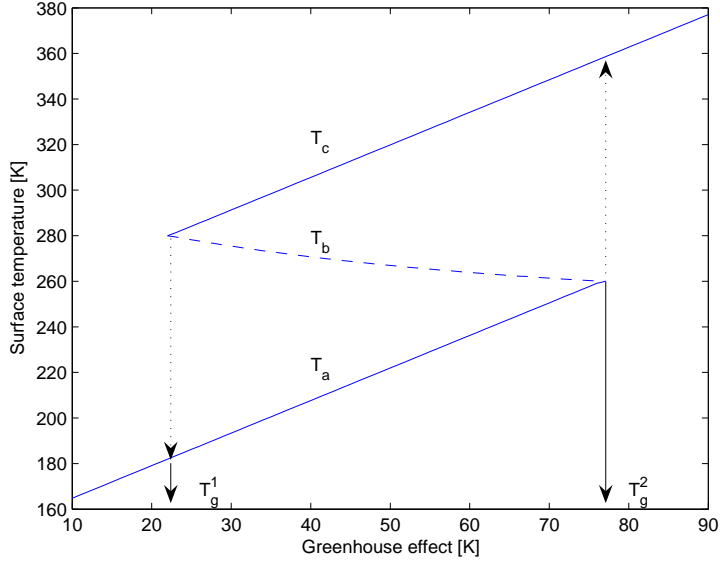}
\end{center}\caption[]{
Equilibrium temperature as a function of CO$_2$ concentration.
Courtesy of P.\ Ditlevsen \cite{Ditlevsen05}.
}\label{ditlevsen2}\end{figure}

Ditlevsen \cite{Ditlevsen05} used \Eq{dTdt} to study the response
of the system to variable greenhouse gas concentrations.
As the amount of CO$_2$ increases, the equilibrium temperature
increases.
Obviously, when the system was on the lower fixed point initially,
there must be a critical CO$_2$ concentration above which
the solution will jump discontinuously to the upper branch;
see \Fig{ditlevsen2}.

\begin{figure}[t!]\begin{center}
\includegraphics[width=.9\columnwidth]{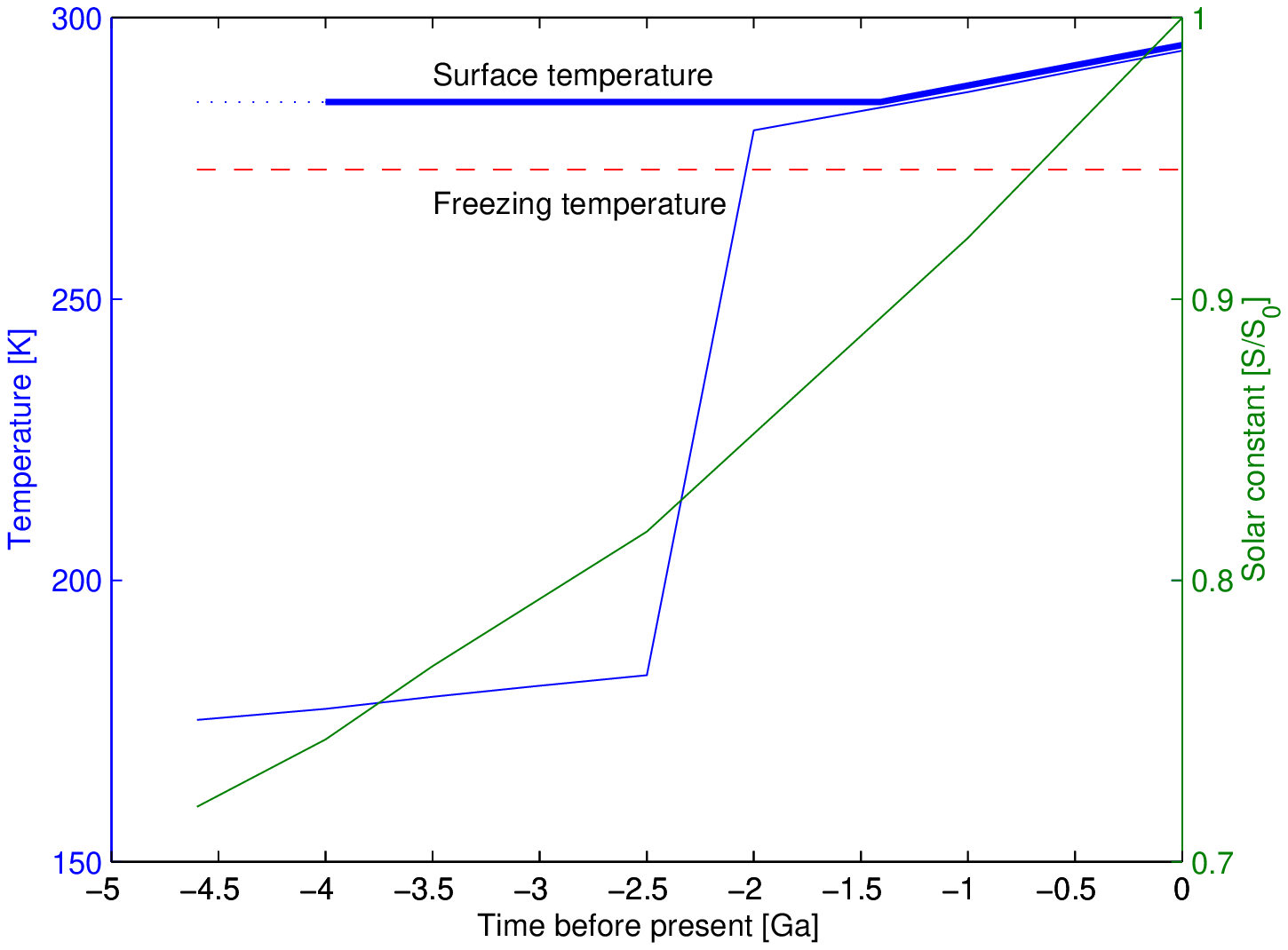}
\end{center}\caption[]{
Dependence of surface temperature on time under the assumption
of a continuous source of outgassing of CO$_2$ and the onset of
a CO$_2$ sink for $T>T_{\rm w}$.
Courtesy of P.\ Ditlevsen \cite{Ditlevsen05}.
}\label{ditlevsen3}\end{figure}

It is generally accepted that the rate of weathering increases with
increasing temperature.
This provides a stabilizing effect on the climate.
As $T$ increases, the rate of weathering increases, removing more CO$_2$
from the atmosphere, reducing the greenhouse effect, and thus leading
to cooling.
Ditlevsen \cite{Ditlevsen05} introduced the assumption
that there is a continuous source of CO$_2$
through outgassing from volcanoes and a temperature-dependent sink
of CO$_2$ from weathering when $T$ exceeds a critical temperature
$T_{\rm w}$, but no weathering for $T<T_{\rm w}$ \cite{Walker}.
This leads to a self-regulating effect for $T<T_{\rm w}$, which
Ditlevsen calls a greenhouse thermostat.
Whenever $T<T_{\rm w}$, since there is then no weathering and hence
no sink of CO$_2$, greenhouse gases will build up until the Earth's
temperature has reached the value $T_{\rm w}$; see \Fig{ditlevsen3}.
This is the mechanism that is believed to have caused the early
Earth to be above freezing through most of its history---with the
exception of intermediate Snowball Earth-like events that are
caused by the emergence of other sinks of greenhouse gases such
as the onset of aerobic photosynthesis or the enhanced formation
of mountain topography that leads to an increase in the erosion
rate and hence the weathering.

\subsection{The Daisyworld model}

Life can also affect the planet's albedo, as has been demonstrated by
Lovelock \cite{WatsonLovelock83} in his Daisyworld model.
For a tutorial on the Daisyworld model see Ref.~\cite{vonBloh}.
This model also makes use of \Eq{dTdt}, but now 
the planet's albedo $A$ is affected by the plant
population which is simplistically represented by black and white plants
or flowers (daisies) with local albedos $A_1$ and $A_2$, respectively.
So the total albedo is a weighted average of the form
\EQ
A=\sum_{i=1}^3\alpha_i A_i,
\EN
where $A_3$ is the albedo of the unpopulated surface.
The weights $\alpha_i$ depend on the surface coverage of the
respective regions and obey evolution equations that are in turn
governed by a temperature-dependent growth term, $\beta(T_i)$, and a fixed
death rate, $\gamma$, so the resulting equations for the rate of change
of the albedo are
\EQ
{\dd\alpha_i\over\dd t}=[\beta(T_i)-\gamma]\alpha_i,
\EN
where $i=1$ for black and $i=2$ for white plant populations,
$\beta(T)$ is assumed to be different from zero in
the range $T_{\min}<T<T_{\max}$ with a maximum at
$T_{\rm aver}=\half(T_{\min}+T_{\max})$.
The weight for the unpopulated surface follows from the
normalization $\sum\alpha_i=1$, so $\alpha_3=1-\alpha_1-\alpha_2$.

\begin{figure}[t!]\begin{center}
\includegraphics[width=.7\columnwidth]{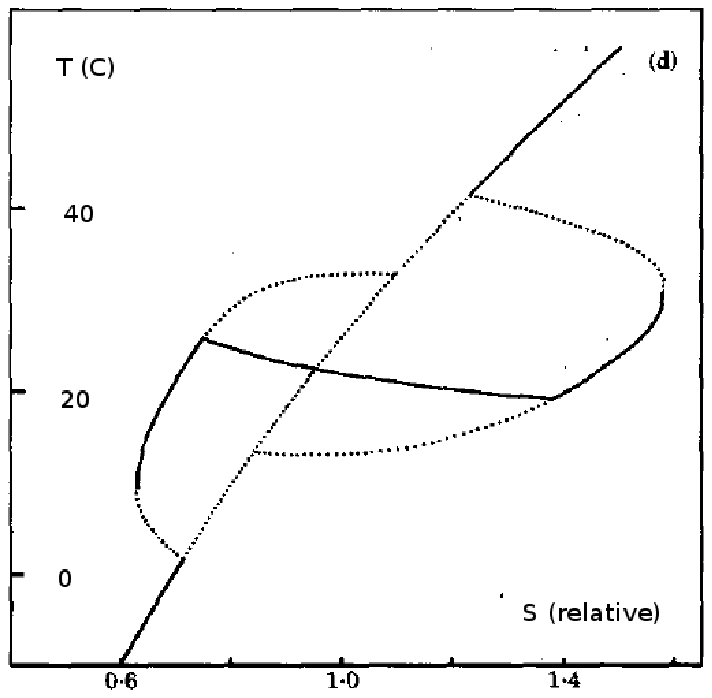}
\end{center}\caption[]{
Temperature (in Celsius) versus relative irradiation (normalized to
the average temperature).
Note that the temperature is stabilized around the value $T_{\rm aver}$,
provided the energy input $S$ is within a certain range.
Courtesy of P.T.\ Saunders \cite{Saunders94}.
}\label{saunders}\end{figure}

The temperatures are higher in the regions of black plants and
lower in regions of white plants according to the formula
\EQ
T_i^4=(A-A_i)q+T_0^4,
\label{qrelation}
\EN
where $q$ is a parameter that must be smaller than a critical value,
\EQ
q<q_{\rm crit}=S/(4\sigma_{\rm SB}),
\EN
in order that heat flows against the temperature gradient \cite{Saunders94}.
For $q=q_{\rm crit}$ the temperature is uniform for different values of $A_i$,
while for $q<q_{\rm crit}$ the regions of high albedo are cooler and
those of low albedo warmer.
Note also that \Eq{qrelation} preserves heat balance, i.e.\
$\sum\alpha_i T_i^4=T_0^4$.

The important point in the Daisyworld model is the fact that, for a
certain range of $S$, the surface temperature of the planet, $T_0$,
is stabilized to be in a certain temperature range around the optimal
value close to $T_{\rm aver}$; see \Fig{saunders}.

Saunders makes another remarkable point.
He showed that by changing the model to allow for Darwinian evolution in
such a way that each plant species works with an optimized temperature
dependence, so $\beta(T_i)\to\beta_i(T_i)$ is modified to become
dependent on $i$, the overall result changes only very little.
More importantly, the range over which the model can stabilize the
planet's temperature shrinks, making the planet as a whole more vulnerable.
Although the amount of shrinkage is small, it emphasizes the dangers
associated with adopting changes that only lead to short-term benefits.
Saunders emphasizes in his work that the ability of life to regulate the
surface temperature of a planet is not associated with natural selection
as in the concept of Darwinian evolution.
More generally he warns therefore that not everything that is to an
advantage needs to be the result of natural selection \cite{Saunders94}.

\subsection{Oxidation of the Earth's crust}

It has recently been proposed that, in addition to the effects discussed
above, life may have profound effects also on the Earth's crust.
A possible scenario of such a suggestion has recently been discussed
by Rosing et al.\ \cite{Rosing06}.
The idea is that photosynthetic life may tap large amounts of solar energy
that were used to reduce carbon from $\mbox{CO}_2$ to compounds of the form
$(\mbox{CH}_2\mbox{O})_n$ and similar, via reactions of the form
\EQ
\mbox{CO}_2+\mbox{H}_2\mbox{O}+h\nu
\rightarrow\mbox{CH}_2\mbox{O}+\mbox{O}_2,
\EN
where $h\nu$ denotes the energy taken from solar radiation.
Furthermore, and even more surprisingly, the oxygen produced by
photosynthesis may have been critical in oxidizing iron
in the continental crust.
Although other factors also plaid a substantial role, it is clear that
biological processes can speed up the oxidation process substantially.
Comparing the oceanic crust with continental crust, a major difference
is the enhanced fraction of SiO$_2$ (57\% in the continental crust compared
to 50\% in the oceanic crust).

With granite being one of the lightest rock types, it was eventually
able to escape subduction and to produce stable continents about
$3.8\Gyr$ ago.
This is also the time of the oldest rock findings on Earth.
Given that the rise of continents on the early Earth is associated
with granite formation, the presence of granite on silicon-bearing
rocky planets might thus be a possible biomarker for photosynthesis
\cite{Rosing06}.

Although this idea is speculative, it may be supported quantitatively
as follows.
Firstly, the present day production rate of organically fixated
carbon is estimated to be $9\times10^{15}\mol\,\mbox{C}\yr^{-1}$
\cite{Martin,desMarais}.
The amount of energy required for this can be calculated by using
the fact that it costs $477\kJ$ to transfer one mol carbon to hexose.
The energy required for this is then $300\mW\m^{-2}$.
Rosing et al.\ \cite{Rosing06} argue that this amount could be supplied
by only 0.1\% of the effective solar energy flux, $S/4$.
Assuming that the amount of carbon burial, relevant to estimating
the usable fraction of oxygen for iron oxidation, is also about $0.1\%$,
this corresponds to about $10^{13}\mol\;\mbox{C}\yr^{-1}$.
This would yield a comparable iron oxidation rate.
Rosing et al.\ \cite{Rosing06} argue further that the annual basalt
production contributes about $10^{14}\mol\;\mbox{Fe}\yr^{-1}$, so a
fraction of the magmatic iron flux could be used for building up the
mantle reservoir of ferric iron.

In conclusion, the presence of life can lead to significant alterations
of the planet in a number of different ways,
as is quite clearly demonstrated by some of the differences
between Earth and its neighboring planets Venus and Mars.
Only the Earth has extensive reservoirs of oxygen and of granite.
Within limits, the presence of life on a planet can also have a
stabilizing effect on its climate.
The relevant mathematical modeling of some of these processes
resembles in many ways those
encountered earlier in studies of homochirality and of the spread
of hereditary information on the early Earth.

\section{Conclusions}

Astrobiology has developed into a rapidly growing research field
involving expertise from a number of neighboring disciplines.
Nonlinear dynamics and nonequilibrium thermodynamics find
applications in all these subfields.
Here, we have elaborated on a few such aspects.
Closest to the onset of life is perhaps the emergence of homochirality
of biomolecules.
Given that RNA has been proven to form longer polymers
only in a homochiral environment,
one would expect that homochirality must be a prerequisite to the emergence
of life at the level of a replicating RNA world.
On the other hand, the very mechanism causing the polymerization to
terminate, namely the
enantiomeric cross-inhibition, can also be the mechanism responsible
for causing 100\% homochirality by destroying RNA molecules whose
chirality is already in the minority.
This would however requite the possibility of auto-catalysis, which
can be avoided in another scenario where a closed peptide system
is kept away from equilibrium by continuous activation of amino acids.

Chemically speaking, the stabilization of a definite chirality is in some
models similar to the subsequent establishment of hereditary information
in that catalysis plays a crucial role.
Furthermore, in both cases the possibility of chemistry in an
extended system is crucial.
On the one hand, spatial extent gives rise to the possibility of
coexistence of life forms of opposite handedness on the early Earth.
On the other hand, spatial extent can be critical in allowing the
system to find unpopulated locations faster than being overwhelmed by
the effects of parasites that tap the same resources that are required
for the maintenance and development of hereditary information.

Finally, life is invariably coupled to come kind of metabolism
that is ultimately powered by solar energy.
This clearly affects the environment by reducing carbon and
oxidizing  the crust of the Earth and, over the last two billion
years also the atmosphere.
How much these alterations of the environment are due to biological
processes is less obvious.
However, it is clear that biological factors greatly speed up weathering
on the Earth.
The extent of biologically induced alterations of the continental
crust, for example, may therefore best be tested using quantitatively
accurate model calculations.
The outcome may ultimately hinge on energetic considerations and on
the efficiency of photosynthesis as a solar energy collector.

With the scope of being able to explore in the near future not only the
planets and other celestial bodies in the solar system in much more detail,
but also planets of other planetary systems, the research in
astrobiology quickly develops into a field that will be driven more and
more by new data, making this field less susceptible to speculation.
It is therefore important to be prepared for upcoming discoveries
in this field.
Finally, it should be emphasized that astrobiology is efficient in
communicating science to the general public, which may provide additional
boost to the field.

\section*{Additional reading}

\begin{list}{}{\leftmargin 3em \itemindent -3em\listparindent \itemindent
\itemsep 0pt \parsep 1pt}\item[]

Barbieri M\ybook{2003}{The organic codes: an introduction to semantic
biology}{Cambridge: CUP}

Brack A\ybook{1998}{The molecular origins of life}
{Cambridge University Press, Cambridge}

Darwin C\ybook{1859}{The origin of species by means of natural section}
{Reprinted by Penguin books, London 1985}

Haken H\ybook{1983}{Synergetics -- An Introduction}{Springer, Berlin}

Lovelock JE\ybook{1995}{Gaia. A new look at life on Earth}
{Oxford Univ. Press, Oxford}

Lunine J\ybook{2003}{Astrobiology: a multi-disciplinary approach}
{Pearson Addison-Wesley, San Franciso}

Prigogine I\ybook{1980}
{From being to becoming. Time and complexity in the physical sciences}
{Freeman, New York}

Rauchfu{\ss} H\ybook{2005}{Chemische Evolution und der Ursprung des Lebens}
{Springer, Berlin}

Ward PD, Brownlee D\ybook{2000}{Rare Earth: why complex life is
uncommon in the Universe}{Copernicus, Springer: New York}


\end{list}

\vfill\bigskip\noindent\tiny\begin{verbatim}
$Header: /var/cvs/brandenb/tex/bio/encyc/paper.tex,v 1.39 2007/08/12 21:43:21 brandenb Exp $
\end{verbatim}

\end{document}